\documentclass[runningheads]{llncs}

\usepackage[hyphens]{url}

\usepackage[left=123px,right=123px,top=115px,bottom=110px]{geometry}

\usepackage{amsmath, amssymb,url}
\usepackage{graphicx}
\usepackage{balance}
\usepackage{xcolor}

\usepackage{caption}

\usepackage[linesnumbered,ruled,vlined]{algorithm2e}
\usepackage{subfigure}

\usepackage{comment}

\SetAlFnt{\scriptsize}
\SetAlCapFnt{\footnotesize}
\SetAlCapNameFnt{\footnotesize}

\begin{document}

\title{SocialSpy: Browsing (Supposedly) Hidden Information in Online Social Networks}
\titlerunning{SocialSpy: Browsing (Supposedly) Hidden Information in OSNs}

\author{Andrea Burattin\inst{1}\and Giuseppe Cascavilla\inst{2,3}\and Mauro Conti\inst{1}}

\institute{University of Padova, Italy\\
\email{\{burattin,conti\}@math.unipd.it}
\and
University of L'Aquila, Italy\\
\and
VU University, Netherlands\\
\email{g.cascavilla@student.vu.nl}
}


\maketitle

\begin{abstract}

Online Social Networks are becoming the most important ``places'' where people share information about their lives. With the increasing concern that users have about privacy, most social networks offer ways to control the privacy of the user. Unfortunately, we believe that current privacy settings are not as effective as users might think.

In this paper, we highlight this problem focusing on one of the most popular social networks, Facebook. In particular, we show how easy it is to retrieve information that a user might have set as (and hence thought as) ``private''. As a case study, we focus on retrieving the list of friends for users that did set this information as ``hidden'' (to non-friends). We propose four different strategies to achieve this goal, and we evaluate them. The results of our thorough experiments show the feasibility of our strategies as well as their effectiveness: our approach is able to retrieve a significant percentage of the names of the ``hidden'' friends: i.e., some 25\% on average, and more than 70\% for some users.





\end{abstract}
\section{Introduction}

Online Social Networks (OSNs) are web applications that allow users to build connections and establish relationships to other Internet users. Social networking can be used to stay in touch with friends, make new contacts and find people with similar interests and ideas. These online services have grown in popularity since they were first adopted on a large scale in the late 1990s. Geocities was among the first social networking sites on the Internet, launching its website in 1994 \cite{historySocialNetwork}. 
Currently, one of the most famous and used OSNs is Facebook.
On October 4th, 2012, Facebook reached one billion users \cite{FBstatistics}.
All these users generate, share, and link a huge amount of data. According to statistics as of August 2012 \cite{FBamountDATA}, 300 million photos were uploaded daily, 2.7 billion likes were made daily, and an average of 500 terabytes of new data were stored on a daily basis. Looking at the recent growth rate of Facebook, it is possible to assert that its only limit is given by two factors: the world population and government policies. Specifically, as reported in \cite{1,2,4}, Facebook is still not available for all the users around the world. Facebook, 
in order to handle such growth, went through a lot of improvements, many of them closely related to privacy and security issues \cite{5}. However, as reported for example in \cite{FBsecurity,6195711}, the platform is still affected by data leakages.

We argue that one of the biggest challenges of OSNs is to ensure privacy and security for the data of their users. In \cite{Luo2009}, authors give a generic idea about methods and reasons that push a user to launch an attack against OSNs. In particular, the authors claim that the basic motivation behind such attacks is an economical reward, and most of the users just use ``horse sense'' to protect themselves. Moreover, the same study pinpoints the OSNs as being responsible of warning users against the risk of attacks, enhancing their spam filters, and taking care of application bugs. For these reasons, OSNs privacy and security are well studied issues. 

To ensure the privacy of the data, OSNs give to their users tools to set up rules for the visibility of their data. In Facebook, for example, there is the possibility to divide friends in ``\emph{Folders}'', and set different privacy rules for each folder.  In Google Plus there are ``\emph{Circles}'' to define different privacy rules for the different group of friends. LinkedIn and Twitter do not have custom privacy rules for specific set of contacts.
The main problem with privacy settings is to learn how to properly use and tune them. Several guides, today, are available on Internet, with the explicit goal to help people in such configuration. However, even when the user configures his profile in a proper way, problems may arise with the privacy settings of third-party applications. For example, the current Graphical User Interface (GUI) of Facebook does not help inexperienced users in understanding what kind of permissions are better to give to an application to keep a good level of privacy \cite{10}. Using this Facebook GUI, a user simply authorizes the application to have access to (all) his data. Once the application is authorized, the data from the user could become publicly available.
Almost 13 million users said they had never set, or did not know about Facebook's privacy tools. Furthermore, 28\% share all, or almost all, their wall posts with an audience wider than just their friends \cite{FBprivacy}. According to our studies and experiments, we hardly believe that users are completely aware of actual privacy that OSNs provide them. On the other hand, whenever users know that their profiles have some information leakages, they are often too lazy (or inexperienced) to properly modify the privacy options and make the profile private \cite{Madejski2012}.

The approach we propose in this paper can be considered as Open Source INTelligence (OSINT) techniques. In the intelligence community, the term ``open'' refers to overt, publicly available sources (as opposed to covert or clandestine sources). OSINT approaches aim at extracting knowledge from publicly available sources \cite{OSINT,steele2007open}. In fact, our study, uses only information publicly available on Facebook. In particular, the tools we use in our strategies are:
\emph{(i)} Graph APIs \cite{GraphAPI}, i.e., set of APIs which allows developers to build applications capable of reading and writing user data;
\emph{(ii)} Mutual Content Page \cite{MutualContentPageFB}, i.e., a page that displays which content two users have in common;
\emph{(iii)} FacePile plugin \cite{SocialPlugin}, i.e., a web page that displays a portion of the people who liked a given page. The FacePile plugin is usually used in a website to show who liked the related Facebook fan page.\\                                                                                                                                                                    
We underline that all these tools are freely available to any Facebook user.

The contribution of this paper is threefold. We first underline the problem that Open Source Intelligence (OSINT) \cite{conf/secrypt/KandiasMSG13,PMID:23479631} applied to Online Social Network allows to retrieve a significant amount of information that the user consider, set, and think as remaining private.
Then, we propose four different practical OSINT strategies to retrieve the Friends List from a popular OSN (i.e., Facebook). Finally, we prototype our strategies, and run a set of experiments. Our implementation demonstrates the feasibility of our approaches, while the results of the experiments show their effectiveness: our approach is able to retrieve a significant percentage of Facebook identifiers (IDs) of the ``hidden'' friends: i.e., some 25\% on average, and more than 70\% for some users.

The remaining part of the paper is organized as follows. In Section~\ref{relatedw} we review the state of the art. In Section~\ref{systemModel} we give a formalization of Facebook and of our framework. In Section~\ref{hiddenInfo} we present the weakest part of Facebook. In Section~\ref{strategies} we describe our strategies to retrieve data. In Section~\ref{evaluation} we present our experimental settings and we discuss the results.  Finally, Section~\ref{conclusion} draws some conclusions.

\section{Related Work}
\label{relatedw}

In the literature, there are several studies about privacy in Online Social Network. 
These works revealed the lack of privacy and security in OSNs and how simple it is to get private information about users.
The literature is split between studies on OSNs privacy issues on one hand, and possible data protection solutions on the the other side.
We can then divide this related work section in three parts: ``Attacks'', ``Solutions'' and ``Motivations''. Under the ``Attacks'' section we are going to present all those studies that try to retrieve information, considered private, from OSNs. Instead, in ``Solutions'', we analyze the proposed way to protect users data in OSNs. Lastly, in ``Motivations'', we discuss some issues that make user profiles in OSNs lacking of privacy.

\subsubsection{Attacks}

As part of ``Attacks'' studies we have the work described in \cite{6195711}. It aims at retrieving user age by crawling through Facebook \cite{7}. 
The study shows how to retrieve the victim's age using ``reverse friend lookup technique''. 
Indeed, exploiting the lack of privacy of victim's friends it is possible to retrieve personal information from the profile of the victim itself.
Similarly to the study above mentioned is \cite{Tang:2011:WNS:1996686.1996731}. The study uses a crawling technique to retrieve names of users from Facebook. Cong Tang et al. use for the experiment only public profiles from Facebook and citizens of New York City. The study processed all the properties of the retrieved names and compared them with a popular name list obtained via offline mechanisms. Having the names of users with a public profile in Facebook \cite{Tang:2011:WNS:1996686.1996731} tries to infer the gender of the retrieved users.
Another experiment based on crawling is presented in \cite{Thomas:2010:UMP:1881151.1881165} where Kurt et al. show the failure of privacy options provided by Facebook to protect users from personal content leakage by friends.
An interesting study, that tries to discover hidden information about a victim user is reported in \cite{6542422}. In this work, authors
use public information available from a social network. Collecting the information retrieved from the social network, they build queries for search engines.
With the queries and the corresponding results they try to discover new information about the victim.
Costantino et al. \cite{6567223} proposes Phook, a Facebook application. Once the user authorizes Phook to have access to his data,  the application is able to retrieve, from the profile of the user, photos based on keywords. 
Using Facebook Query Language (FQL), Phook makes queries on the Facebook user profile and retrieves all the 
photos related to the keywords. However, Phook does not discover hidden information from users on Facebook. Phook retrieves information from the profile of the user that is connected to the Phook application.

\subsubsection{Solutions}

If above we have a handful of attacks, the literature proposes also some solutions to mitigate the lack of privacy in OSNs.
An approach is presented in \cite{5283227}, where Luo et al. propose a solution for the 
disclosure of information based on the architecture ``FaceCloak'', where information from a user are hidden to the non authorized profiles.
The study described in \cite{Conti2011} proposes a system which is able to enforce privacy protection in Facebook.
With a Firefox plugin, called FaceVPSN, Conti et al. try to mitigate the problem of lack of privacy in Facebook. 
FaceVPSN is completely distributed, Facebook independent, and hides information from users outside the 
Virtual Private Social Network (VPSN). 
This plugin hides public information ``covering'' them with fakes. Then, only those users that are part of the VPSN network can retrieve the real information
of a user. Solutions such as FaceVPSN provides user anonymity via registering a fake identity in Facebook. However, de-anonymization solution have 
also been proposed \cite{5207644}. Hence, solutions as FaceVPSN should be used in conjunction with mechanisms that make the  de-anonymization of networks more hard \cite{6529495,VirtualFriendship}.
Buchegger et al. propose PeerSoN \cite{Buchegger:2009:PPS:1578002.1578010}: a tool to address privacy concerns over Online Social Networks (OSNs). This tool is based on a peer-to-peer architecture and  provides users with privacy protection in OSN. This system is decentralized and independent from any OSN: it is able to encrypt the communication with the OSN.
When a peer wants to connect to another peer, it first queries the look-up service in order to get all required information. Then, peers are able to directly connect each other. Once the peers exchanged a message or file, they immediately disconnect.  This architecture, however, is still a prototype.
An approach to mitigate the problem of Fake Profile Attack in OSNs is showed in \cite{6425616}. 
The aim of this work is to understand, from the behavior of a user in OSNs, if a Facebook profile is real or fake.
In \cite{6425616}, the authors made their experiments looking at some different variables as 
``\emph{Evolution over time of the number of friends}'', ``\emph{Real life social network based verification}'', and
``\emph{OSN graph structure for fake profile detection}''.

\subsubsection{Motivations}

Lastly, as part of motivation part, we try to understand how it is possible this lack of information in OSNs. One of the many reason for the disclosure of information in OSNs is reported in \cite{5231850}. Through a simple experiment, 
Nagle and Singh show how easily people accept a request of friendship on Facebook from a stranger. 
The percentage of people that add an unknown user on Facebook increases if there is a common friend between the user and the unknown new friend.  This is also due to the lack of attention from the user that does not care too much about the problem that someone can steal his information.
%
Beyond the reasons that lead users disclosure information, it is necessary to state that, in recent years, profiles in OSNs are becoming more private.
The experiment, concerning privacy in OSNs reported in \cite{DBLP:conf/percom/DeyJR12}, shows
the raising of awareness from users concerning the lack of privacy in OSNs. Ratan et al. ran the same experiment twice; 
the first one in 2010 and the second one in 
2011 and compared the results. Crawling some profiles, that are part of the NYC subnetwork, they 
discovered that in 2010 the amount of hidden profile was close to 17\%. Ratan et al. restarted the crawler on the same subnetwork in 
2011 and the percentage of hidden profiles has been increased up to 52\%.  
Therefore, according to this paper, users had a growth of awareness and started to take care of their online data and on the OSNs as well.

\vspace{1em}
The work we present in this paper aims at understanding how easily and how much supposedly hidden information can be retrieved from a Facebook profile. In fact, all the above-mentioned studies are related to discover some simple 
information of a given user or to try to protect the information itself. 
Instead, we propose an approach to retrieve information, set as private, from a user Facebook profile. The first part of our study can be classified under the ``Attacks'' field. On the other hand, the study aims to create awareness in OSN users. Through our attacks we show how a user can protect his information using privacy options provided by OSNs. Moreover, we show what privacy options the OSNs require an update in order to better protect its users information. These last part can be classified as ``Solutions''.
Our strategies, differ from the above mentioned studies on several aspects: 
\emph{(i)}~we do not need any type of authorization from the user to have access to his data (our strategies are not based on a Facebook application);
\emph{(ii)}~we keep the victim user completely unaware of the attack;
\emph{(iii)}~we are able to rebuild the friends list of a user that is not a friend of the attacker (they do not have to share any information);
%
\emph{(iv)}~we are able to retrieve the friends list set as ``\textit{private}''; 
\emph{(v)}~we work on random profiles from Facebook, on which we do not have any type of information, except the username;
\emph{(vi)}~neither information nor friends are shared between our experimental profiles and the victim profiles.

\section{System Model}
\label{systemModel}

Facebook is composed of different entities. All these entities, together, give the possibility to the final user to perform 
different actions into such ``ecosystem''. The entities we consider are: \textit{pages}, \textit{users}, \textit{groups} and 
\textit{pictures}. 
\emph{Users} are allowed to perform some actions: become ``friend'' of another user; ``like'' a \emph{page} (and revoke the ``like''), ``join'' a \emph{group} (and leave the group), and ``like'' or ``comment'' pictures (and revoke the ``like'' or delete the ``comment'').
Instead, \textit{pages}, \textit{groups} and \emph{pictures} are ``passive'' entities (i.e., they are managed by \emph{users}).
\textit{Pages} are always public. The set of pages a user likes can be interpreted as the \emph{tastes} of that user. Usually \textit{pages} enable public figures (such as companies, organizations, or celebrities) to create a presence on Facebook \cite{defLikeGroup}.
\textit{Groups} on Facebook are ``places'' where people can share and discuss their common interests and express their opinion around common causes, issues or activities to organize \cite{defLikeGroup}. A group is not always public: tuning its privacy rules, it is possible to set it as public (accessible and searchable to all users in Facebook), private (accessible only if invited; searchable) or hidden (accessible only if invited; not searchable).
\textit{Pictures} are usually uploaded by users. On Facebook, it is really difficult to take under control the privacy settings
of pictures. There are pictures directly uploaded by a user, pictures where users are tagged, cover photos (that are always public) and profile pictures.

More formally, the portion of Facebook that we are going to use in the rest of this paper can be formalized as the tuple:
$
	\textit{Facebook} = (\mathbb{P}, \mathbb{U}, \mathbb{G}, \mathbb{I}).
$
Specifically, in this notation, we have that:
\begin{itemize}
	\item $\mathbb{P}$ is the set of pages. A page $p \in \mathbb{P}$ is something related to the tastes of a user, i.e., what a user might like. 
	\item $\mathbb{U}$ is the set of users. A user $u \in \mathbb{U}$ represents a person. Each person can ``like'' a page $p$, join
	a group $g$, leave comments into a page, request friendships to other users (accept friendship from other users), 
	upload pictures into his own profile pages.
	\item $\mathbb{G}$ = $(G', n)$ is the multiset that represents groups, where $G' \subseteq \mathcal{P}(\mathbb{U})$ (given a set $A$, $\mathcal{P}(A)$ is the power set of $A$, i.e., the set of all subsets of $A$) and $n: G' \to \mathbb{N}_{\geq 1}$ is the 
	multiplicity function. $\mathbb{G}$ represents all the groups on Facebook (please note the same set of users may appears several times).
	A group, from the ``application'' point of view, is a place where a user can promote, share and discuss relevant topics.
	\item $\mathbb{I}$ is the set of pictures. Every picture $i \in \mathbb{I}$ can receive one or more ``likes'' and one or more ``comments''  from a user $u \in \mathbb{U}$. Therefore, it is possible to consider a picture as the pair $i = ({U^l_i}, {U^c_i})$. Where ${U^l_i} \subseteq \mathcal{P}(\mathbb{U})$ is the set of users that liked $i$, and ${U^c_i} \subseteq \mathcal{P}(\mathbb{U})$ is the set of users that commented on $i$.
\end{itemize}

Within our model, a user $u$ is defined as the tuple
$
	u = (\textit{Personal}, U, P, G, I),
$
where: $\textit{Personal}$ is the set of ``personal''  information (such as the name, the family name, the age), $U \subseteq \mathbb{U}$ is the set of friends of $u$; $P \subseteq \mathbb{P}$ is set of pages $u$ likes; 
$G \subseteq \mathbb{G}$ is the set of groups $u$ belongs to; and $I \subseteq \mathbb{I}$ is the set of personal pictures (pictures that $u$ uploaded into the social network).

Due to all these interacting entities, and their complex set of privacy settings, it is very easy to observe information leakages out of Facebook.

\section{Retrieving Hidden Information}
\label{hiddenInfo}

The aim of this work is to retrieve the lists of friends of a victim.  We decided to have as target the friends list because we believe this is one of the most important information on Facebook. In fact, this list might be interesting as a starting point to find even more information. Specifically, from these friends, it is possible to retrieve pictures or comments where the victim user is tagged, or information posted directly from him. It is also possible to consider the revealing of friends list as the \emph{foundation stone} of the process of rebuilding an entire victim profile.

To retrieve the friends list, we start looking at which information can be exploited from the profile of the victim. Specifically, it is possible to categorize the information to exploit in:
\begin{itemize}
	\item Personal Info: general information about the user, which exist independently from Facebook. Examples of this type of information are the gender, the address, the job, the hometown, and the phone number.
	\item Facebook Related Info: information which exist because of the profile page on Facebook. We can categorize this information in three groups: the \textit{pages} a user likes ($P$), the \textit{groups} ($G$), and \textit{pictures} ($I$).
	\item Application Related Info: information connected to the Facebook applications that the victim uses. For example, applications like TripAdvisor \cite{tripadvisor} 
	could publish data, such as the visited places. Other applications may publish tastes information, such as Spotify \cite{spotify}.
\end{itemize}
The easiest retrievable information, from the profile of user, are the Facebook Related Info: groups, pictures and pages a user likes. 
After some studies we figured out that this information is frequently left public, either on purpose by the users, or by the lack of a proper Facebook privacy settings configuration.
Let us briefly analyze how we are going to exploit this Facebook Related Info:
\begin{itemize}
	\item Groups (${G}$). Using the group pages that belong to the victim user $v$, we find user IDs that are friends of $v$. To do this, we apply two different strategies:
	\begin{enumerate}
		\item Using the Graph API, we retrieve all the groups from the profile of $v$. We sort the groups from the smallest to the largest (w.r.t. their number of subscribers). For each group we retrieve all the user profiles and  validate the friendship with $v$, using Mutual Content Page. This verification is made using the text ``Are friends since (date)''. If the Mutual Content Page shows ``Are friends since (date)'', then we are sure that $v$ and the retrieved user are friends. There is no need to have the friendship with the victim, because the Mutual Content Page is publicly accessible.
		\item Using the Graph API, we retrieve all the groups from the profile of $v$. We sort the groups from the largest to the smallest. Group by group we retrieve all the user IDs and we validate the friendship with $v$ using Mutual Content Page.
	\end{enumerate}
	We did not implement the strategies above illustrated because of the huge number of subscribers that usually are part of a group, and also because not all groups are publicly available. As previously outlined, groups have settings to make them ``private'' and ``hidden''. With the ``private'' setting, the group is not accessible. With the ``hidden'' setting it is not possible to find the group on Facebook.
	\item Public pictures of victim user $v$ (${I}$). Very often it is possible to find public pictures on a Facebook \emph{wall} (the profile main page). After some studies, we figured out that among all the pictures, cover photos are always publicly available and it is not possible to set them private. The strategy exploits the likes and the comments that each picture receives. In particular, given a picture belonging to the victim $v$, we can retrieve all the user IDs that liked or commented the picture. Each of them, using Mutual Content Page, is checked for his friendship with $v$.
	\item Liked pages (${P}$). When a victim user $v$ clicks on the ``Like button'' of a page $p$, he becomes fan of $p$. These information are usually left public on the wall of $v$. With our strategy we retrieve the user IDs that clicked the Like button of the page $p$. All the user IDs of this set are checked for their friendship with $v$, using the Mutual Content Page.
\end{itemize}

\section{Strategies}
\label{strategies}

We are now going to analyze in detail our strategies.
The first three strategies are based on liked pages. We decided to use liked pages, for our strategies, because of the meaning of page itself in Facebook. Pages are made to aggregate people around a common interest. The page could be related to an interest really big, like a page of famous actor. Moreover, a fan page could be related to something really small like a little household music group.
Statistics say that Facebook Like or Share buttons are viewed more than 22 billion times a day.
Furthermore, 7.5 million is the number of sites that contain Facebook Like or Share button \cite{LikeShareStatistics}. From this information, we believe it is really common that users are attracted from friend's interests and attract friends sharing and liking pages. Therefore, we think that the possibility to find friends of a victim, between users that liked a page, is really high. The fourth strategy works on likes and comments of pictures of a victim.

\subsection{Strategy~1 (\textit{S1}): Likes Random Order} 

Working on like pages, \textit{S1} is able to target and retrieve the likes from a victim $v$. Once the strategy has a list of likes, there is a high probability to find a handful of friends. The list of user IDs from like pages, sharing the same interest of our victim, is only the starting point to retrieve the friends list of $v$.
Algorithm~\ref{alg:1} illustrates \textit{S1}. 

\begin{algorithm}[h]
	\DontPrintSemicolon
	\KwData{Victim user $v$}
	\KwResult{Set of friends $U$ of $v$}
	\BlankLine
	
	$P \gets \text{set of pages $v$ likes}$\; 							  \label{1st:line:retrieve(likes)}
	$\textit{CandidateFriends} \gets \emptyset$ \;							
	\ForEach{$p \in P$}{ 										  \label{2nd:line:forech_p_from_P}
		\tcc{Find page fans with FacePile}
		$\textit{CandidateFriends} \gets \textit{CandidateFriends}\ \cup\ \textrm{FanOf}(p)$\; \label{3rd:line:candidate_friends}
	}
	\BlankLine
	$\textit{FriendsFound} \gets \emptyset$ \;
	\ForEach {$c \in \textit{CandidateFriends}$}{							  \label{4th:line:foreach_u_like_p}
		\tcc{Check friendship with Mutual Content Page}
		\If{$\text{AreFriends}(c, v)$ 								  \label{5th:line:check_u_friend_UserID}} {
			$\textit{FriendsFound} \gets \textit{FriendsFound}\ \cup\ \{ c \}$ \;		  \label{6th:line:FriendsFound=FriendsFoundU}
		}
	}  
	\Return $\textit{FriendsFound}$ \;
	\BlankLine
	\caption{Strategy~1 (\textit{S1}), Likes Random Order.}  \label{alg:1}
\end{algorithm}

As mentioned in Section~\ref{systemModel}, this strategy uses $P$ to build $U$. Using the public Facebook page of $v$, \textit{S1} retrieves the liked pages $P$ left public from the profile of $v$ (line~\ref{1st:line:retrieve(likes)} of Algorithm~\ref{alg:1}). The strategy then retrieves the list of fans (the procedure is wrapped on the ``FanOf'' procedure, mentioned in line~\ref{3rd:line:candidate_friends}). To perform this retrieval we use FacePile plugin. Starting from the list of fans, \textit{S1} checks, for each fan if he is a friend of $v$ (line~\ref{5th:line:check_u_friend_UserID}). To understand if two users are friends we use the Mutual Content Page that shows mutual content between two users. If \textit{S1} finds a friend of $v$ it adds the ID of friend in a list called \emph{FriendsFound} (line~\ref{6th:line:FriendsFound=FriendsFoundU}). \textit{S1} repeats this procedure until there are no more candidate friends (friends that like the same pages of $v$).

\subsection{Strategy~2 (\textit{S2}): Likes Ascending Order}

The second strategy is similar to the previous one: once all liked pages from the victim $v$ are retrieved,  \textit{S2} sorts them from the one with the smallest number of fans to the one with the largest. This strategy is designed to start from the page with the smallest number of fans because we assume that these pages are closely related to the interests of our victim $v$, and therefore the probability to find user IDs of friends of $v$ is higher. Once we have some user IDs of  friends of $v$, we can retrieve the list of common friends using the Mutual Content Page itself.
Algorithm~\ref{alg:2} illustrates \textit{S2}.

\begin{algorithm}[h]
	\DontPrintSemicolon
	\KwData{Victim user $v$}
	\KwResult{Set of friends $U$ of $v$}
	\BlankLine
	$P \gets \text{set of pages $v$ likes}$\; 							\label{1st:line:retrieve(likes)}
	$\textit{CandidateFriends} \gets \text{empty priority queue}$ \;
	\ForEach{$p \in P$}{ 										\label{2nd:line:forech_p_from_P}
		$\textit{Fans} \gets \textrm{FanOf}(p)$ \tcc*{Find fans with FacePile} \label{alg:2:GIUSEPPE-DEVI-FARE-TUTTI-I-FIX-CHE-TI-INDICHIAMO}
		$\text{Insert}(\textit{CandidateFriends}, \textit{Fans}, \text{TotalFansOf}(p))$ \tcc*{Inserts the set \emph{Fans} into the \emph{CandidateFriends} queue, with the total number of fans as priority}
	}
	\BlankLine
	Set priority of $\textit{CandidateFriends}$ as min-to-max \;					\label{3rd:line:order_MinToMax}
	\BlankLine
	$\textit{FriendsFound} \gets \emptyset$ \;
	\Repeat{\textit{CandidateFriends} \emph{\textbf{is}} empty} { 				
		$c \gets \text{ExtractFirst}(\textit{CandidateFriends})$ \tcc*{Returns (and removes) the first element of the queue} \label{4th:line:foreachFan}
		\tcc{Check friendship with Mutual Content Page}
		\If{$\text{AreFriends}(c, v)$								\label{5th:line:check_uV_friends}} {
			$\textit{FriendsFound} \gets \textit{FriendsFound}\ \cup\ \{ u \}$ \;		\label{6th:line:FriendsFound=FriendsFoundU2}
		}
	}  
	\Return $\textit{FriendsFound}$ \;
	\BlankLine
	\caption{Strategy~2 (\textit{S2}), Likes Ascending Order.} \label{alg:2}
\end{algorithm}

According to Section~\ref{systemModel}, \textit{S2} uses $P$ to build $U$. Using the public Facebook page of $v$,  \textit{S2} retrieves the liked pages $P$ from the profile of $v$ (line~\ref{1st:line:retrieve(likes)} of Algorithm~\ref{alg:2}). Since \textit{S2} has the set $P$, it organizes them in a priority queue and retrieves the list of fans $U$ from the page using FacePile plugin (line~\ref{alg:2:GIUSEPPE-DEVI-FARE-TUTTI-I-FIX-CHE-TI-INDICHIAMO}).  The liked pages are organized according to a priority min-to-max (line~\ref{3rd:line:order_MinToMax}), it means that Algorithm~\ref{alg:2} tries to find user IDs of friends starting from the page with the less number of fans till the one with the major number of fans. After that, \textit{S2} iterates over the list of fan \emph{CandidateFriends} and checks, for each user ID, if it is a friend of $v$ (line~\ref{5th:line:check_uV_friends}). If this is the case, it adds the user ID in a list called \emph{FriendsFound} (line~\ref{6th:line:FriendsFound=FriendsFoundU2}).
\textit{S2} repeats this procedure since there are no more candidate friends.

\subsection{Strategy 3 (\textit{S3}): Likes Descending Order}

This strategy is very similar to \textit{S2}: we organize the pages starting from the one with the highest number of fans to the smallest (Set priority max-to-min). The main idea behind this strategy is that, fetching like pages from max-to-min, we have the biggest datasets at the beginning of our research where we could find, at least one user ID of friends of $v$. Having one user ID of a friend of $v$ we can use the Mutual Content Page to discover if they have other friends in common and then add the user ID of them to our list of ``Friends Found''.

The Algorithm of \textit{S3} differs from \textit{S2} only at line~\ref{3rd:line:order_MinToMax}: the priority, in the case of \textit{S3}, is set as ``max-to-min''.

\subsection{Strategy~4 (\textit{S4}): Likes and Comments from Pictures}

\textit{S4} does not work with liked pages as Strategies~1, 2 or 3. Instead, it tries to take advantages from pictures of victim profile. The main idea is to use pictures that are left public from victim $v$ or are not settable as private. There are images like ``cover photos'' that the user cannot hide, and are always publicly available. From public images and cover photos, \textit{S4}, retrieves the user IDs of people that pressed the ``Like'' button or commented them. Once \textit{S4} obtains the list of user IDs, checks the friendship between them and $v$ using the Mutual Content Page. Algorithm~\ref{alg:3} illustrates \textit{S4}.

\begin{algorithm}[h]
	\DontPrintSemicolon
	\KwData{Victim user $v$}
	\KwResult{Set of friends of $v$}
	\BlankLine
	$I \gets \text{set of public images of $v$}$\; 							\label{1st:line:PhotosOf(V)}
	$\textit{CandidateFriends} \gets \emptyset$ \;
	\ForEach{$i \in I$}{
		\tcc{Add candidate friends set all users that liked or commented the image}
		$\textit{CandidateFriends} \gets \textit{CandidateFriends}\ \cup\ U^l_i\ \cup\ U^c_i$ \; 	\label{2nd:line:retrieve_UserLikeCommented}
	}
	\BlankLine
	$\textit{FriendsFound} \gets \emptyset$ \;
	\ForEach {$c \in \textit{CandidateFriends}$								\label{4th:line:foreach_u_like_p}} {
		\tcc{Check friendship with Mutual Content Page}
		\If{$\text{AreFriends}(c, v)$									\label{5th:line:check_u_friend_V}} {
			$\textit{FriendsFound} \gets \textit{FriendsFound}\ \cup\ \{ c \}$ \;			\label{6th:line:FriendsFound=FriendsFound+u}
		}
	}  
	\Return $\textit{FriendsFound}$ \;
	\BlankLine
	\caption{Strategy~4 (\textit{S4}), Likes and Comments.} \label{alg:3}
\end{algorithm}

According to Section~\ref{systemModel}, \textit{S4} uses $\textit{I}$ to build $U$. \textit{S4} starts fetching the pictures of victim $v$ (line~\ref{1st:line:PhotosOf(V)} of Algorithm~\ref{alg:3}). From these images the strategy collects all the candidate user IDs  that liked or commented the picture itself (line~\ref{2nd:line:retrieve_UserLikeCommented}).  Once that \textit{S4} obtains this list of user IDs, it checks the friendship with $v$ (line~\ref{5th:line:check_u_friend_V}) using the Mutual Content Page. If \textit{S4} finds a user ID of a friend, it adds it in the \emph{FriendsFound} set (line~\ref{6th:line:FriendsFound=FriendsFound+u}).

\section{Evaluation}
\label{evaluation}

To evaluate our approaches, we conducted some experiments on different datasets. The datasets contain real profiles from Facebook users.

To perform our tests, we logged into Facebook using nine different accounts. These accounts do not have any friends, therefore they are as far away as possible from victims (in terms of ``\emph{hops}'' on the graphs that connects users according to their friendships). The only activity of these accounts consists in liking few pages. 
%
In order not to show an anomalous behavior, that could be detected as malicious by Facebook, we decided to split the load of our requests on nine different accounts. 
With nine accounts we appear like nine different users that make requests to the Facebook servers, each making requests with a lower frequency than the original one. After logging in Facebook with these accounts, we ran our tests to fetch data of victims.


For our experiments we decided to use two types of Facebook datasets: Mixed Dataset and Public Dataset.

\subsubsection{Mixed Dataset}

Using Firefox, we downloaded user IDs from public group pages.
The groups we used are: Universit\'a degli Studi di Padova\footnote[1]{\url{http://www.facebook.com/groups/unipd}}, Roma Giurisprudenza La Sapienza\footnote[2]{\url{http://www.facebook.com/groups/7795542586}}, Studenti dell' Universit\'a degli Studi di Milano - Unimi\footnote[3]{\url{http://www.facebook.com/groups/2310323055}}, Programmers in Padua\footnote[4]{\url{http://www.facebook.com/groups/programmersinpadua}}, F.I.U.P\footnote[5]{\url{http://www.facebook.com/groups/fiupd}}, Pensionati della Polizia di Stato\footnote[6]{\url{http://www.facebook.com/groups/58159395664}}, Flowers\footnote[7]{\url{http://www.facebook.com/groups/246910381993344}}, New York Italians\footnote[8]{\url{http://www.facebook.com/groups/159253824092728}}, Team Ferrari Challange ``Black Jack Caf\'e'' \& Friends\footnote[9]{\url{http://www.facebook.com/groups/WWW.BLACKJACKCAFE}}, The Real Housewives of New York City\footnote[10]{\url{http://www.facebook.com/groups/rhonyc}}.
Most of group pages allow to see their members and therefore it is easy to download their user IDs. This dataset contains public profiles and private profiles, therefore we call it \emph{Mixed Dataset}. Public profiles are composed by those users that decided to have all the content, from their profile, publicly available. There are then private profiles composed by those users that care about their privacy. In this second case the content of their profile is in part or not accessible at all.
This dataset contains 115 
users.

\subsubsection{Public Dataset}

This dataset is public and generated ``around July 15, 2010, by Ron Bowes''. This set of user IDs is publicly available\footnote[11]{\url{http://www.skullsecurity.org}}. Bownes simply generated a tool to download all the Facebook user IDs that are public at the page \url{https://www.facebook.com/directory/}. This page points to all those user IDs that decided to have a public profile on Facebook  or that have answered to the privacy options \textit{``Who can see my stuff?''}, \textit{``Who can contact me?''} and \textit{``Who can look me up?''} with \textit{``Public''} and \textit{``Everyone''}. This public dataset contains 1000 public user names.


\subsection{Experimental Results on Mixed Dataset}
For this and the next paragraph, for readability purpose, all graphs do report a subset of the nodes for which we run experiments.

\begin{figure}[h!]
	\subfigure[Average number of user IDs of friends found with respect to the number of HTTP requests, on Mixed Dataset. \label{fig:stratREQ}]{\includegraphics[width=.47\textwidth]{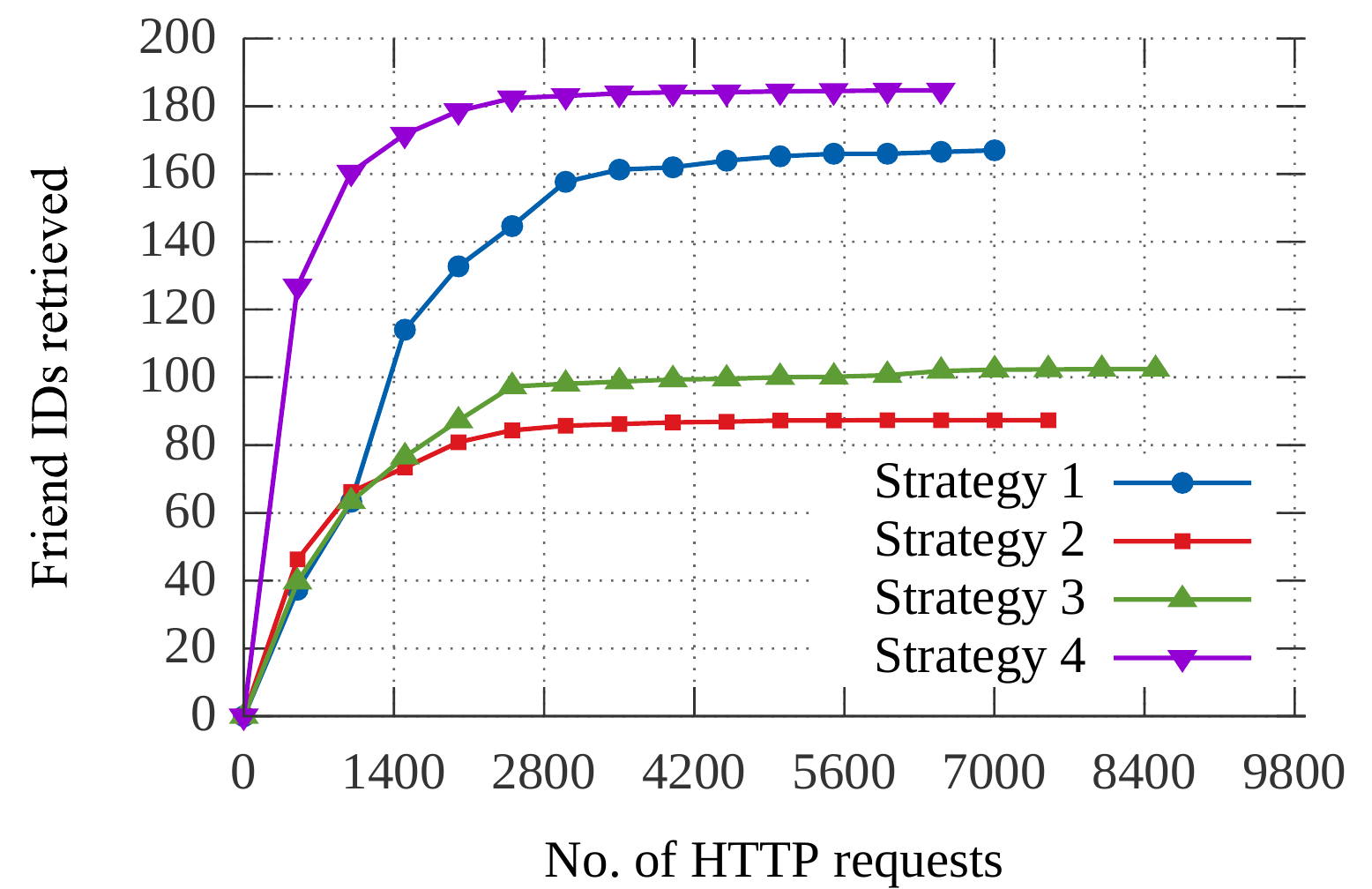}}
	\hfill
	\subfigure[Average number of user IDs of friends found against the time, on Mixed Dataset. \label{fig:stratTIME}]{\includegraphics[width=.47\textwidth]{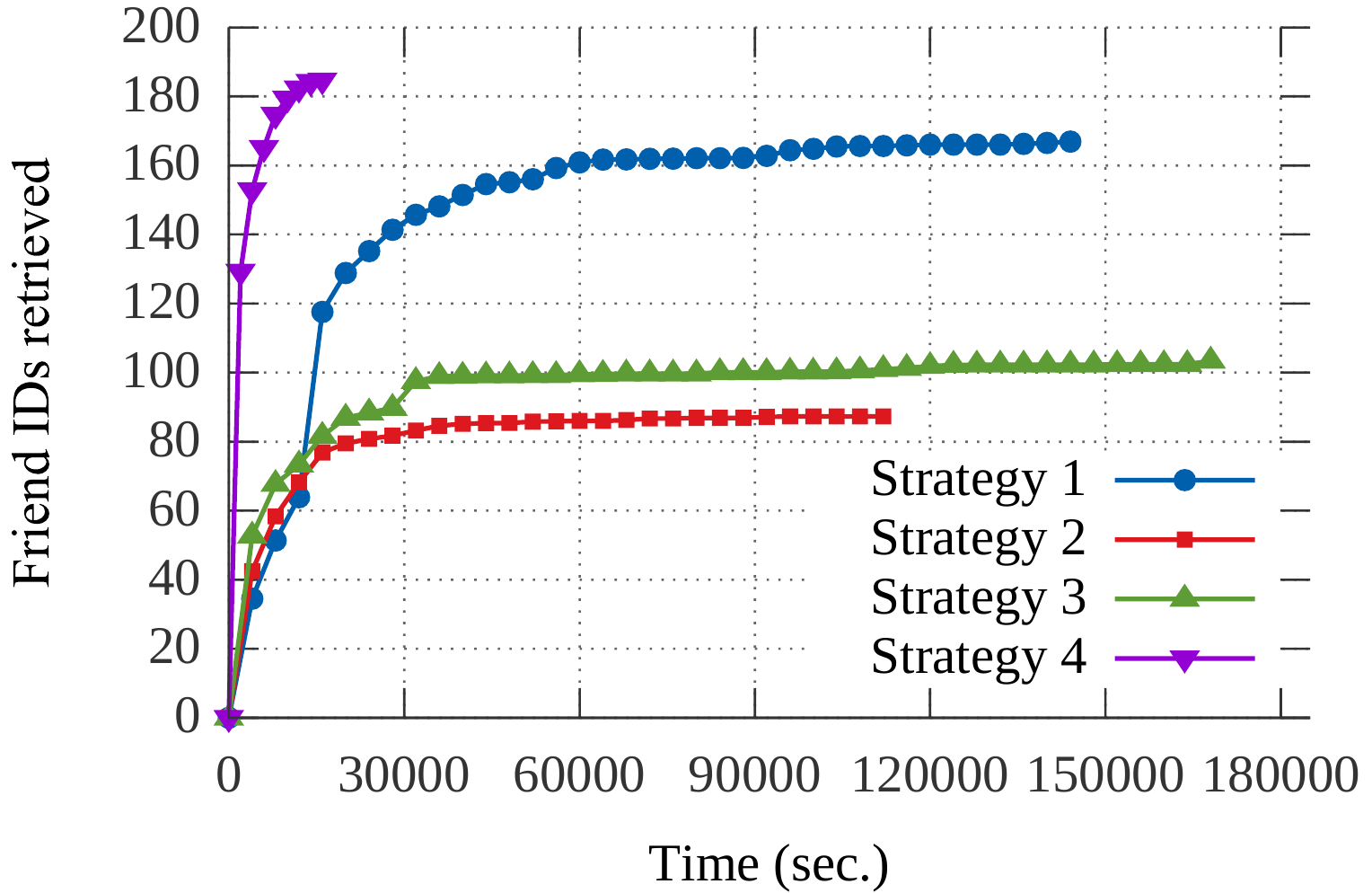}}

	\subfigure[Impact of each strategy in terms of number of user IDs that we succeeded to retrieve data from (on Mixed Dataset. \label{fig:NumberAndPercentage}]{\includegraphics[width=.47\textwidth]{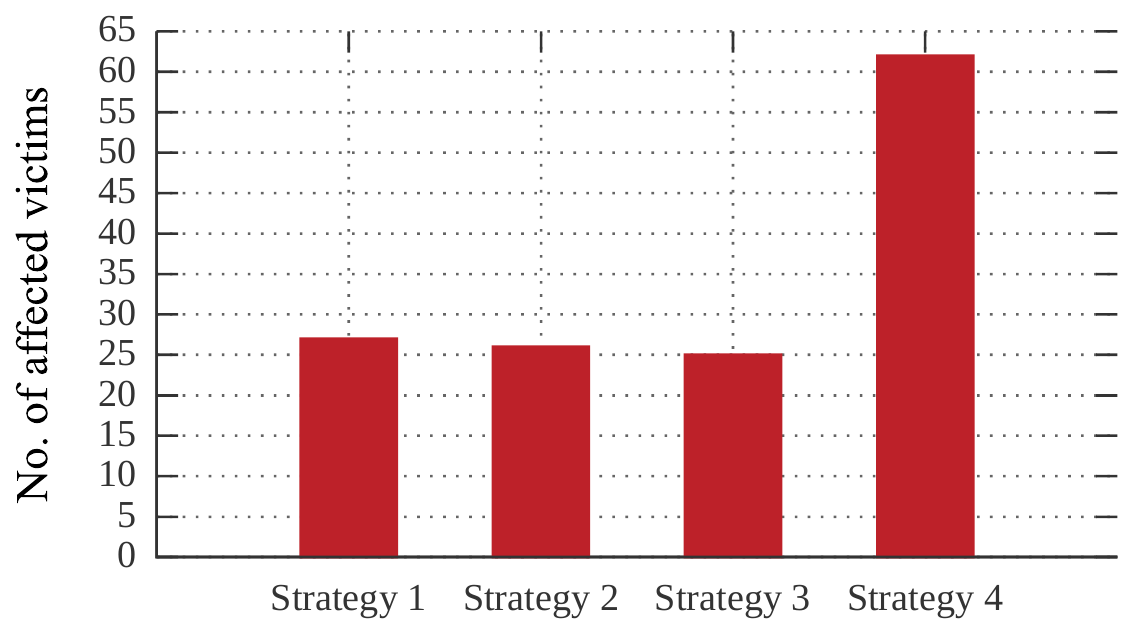}}
	\hfill
	\subfigure[Percentages of user IDs of friends found by each strategy for some users from our Mixed Dataset. \label{fig:FFuser}]{\includegraphics[width=.47\textwidth]{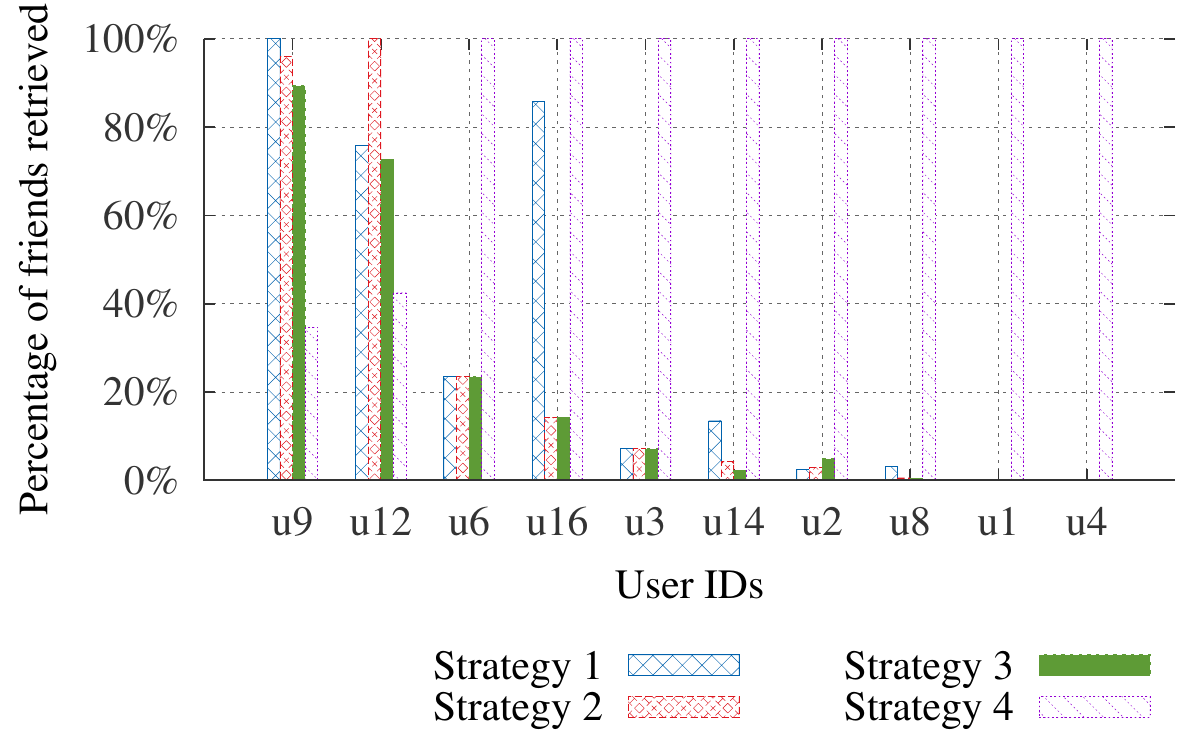}}
	\caption{Performance on the Mixed Dataset.}
	\label{fig:wrapper:mixed}
\end{figure}

Let's first analyze the data coming from the Mixed Dataset.
%
The graph in Fig.~\ref{fig:stratREQ} shows the average of our four strategies by requests on 
115 victims. The $x$-axis shows number of HTTP requests that every strategy requires to retrieve information from Facebook pages. On the $y$-axis, instead, we have the number of user IDs of friends, found by the specific strategy.


Fig.~\ref{fig:stratREQ} demonstrates that \textit{S4} works faster and finds the highest number of friends' user IDs. This is due to the fact that \textit{S4} works on victim's pictures. Since that \textit{S4} uses victim's images, we can assert that most of the comments and likes are from friends of the victim user. It is also possible that non-friend users like or comment victim's pictures. This is due to the fact that public pictures can be available to all the users in Facebook that could decide to like or leave a comment. To recognize who is a friend and who is not, we use the Mutual Content Page given directly by Facebook.

Going deeper into the graph it is possible to note that the four strategies work in a similar way: all the four curves start growing fast at the beginning going to be flat at the end (i.e., they have the same coarse-grained shape). This is due to the fact that all the friends' user IDs are almost found at the beginning and when the strategies do not have more chances to discover friends, the curve becomes flat. 
%
In Fig.~\ref{fig:stratTIME} we draw the graph considering the time (in seconds) to process each request. From the graph below, on the x-axis, it is possible to see how long each strategy worked to find friends' IDs. On the y-axis, instead, how many user IDs of friends every strategy found. \textit{S4} fetches the major number of user IDs of friends.

 

The graph in Fig.~\ref{fig:NumberAndPercentage} shows how many victims our strategies succeeded to retrieve user IDs of friends from their supposed hidden data.
%
%
Let us consider \textit{S4} as an illustrative example: in the graph in Fig.~\ref{fig:NumberAndPercentage}, \textit{S4} retrieved user IDs from about 65 users out of a dataset of 115.  
%
Fig.~\ref{fig:FFuser} shows which strategy works better on some randomly selected victims of the Mixed Dataset. This figure does not show all the victims with the related best strategy, instead it reports just a small view of the performance on few elements of our dataset. Since we are not able to know how many friends our victims have, we decided to consider the strategy that fetched the highest number of user IDs of friends as 100\%. Then we draw all the other strategies around the highest one calculating the percentage as the ratio between: the number of friends found from strategy and the highest number of friends found. Using this graph, we have the possibility to analyze, for every single user, which strategy is the best in retrieving user IDs of friends and the percentage of friends found from the other strategies compared to the best one. Due to the results, we can state that the \textit{S4} obtains better results for most of the users. There are two users, i.e., \textit{u9} and \textit{u12}, where \
\textit{S1} and 
\textit{S2} work better than others. From this graph we can assert that 
\textit{S4} is not always the best one. This is due to the fact that there are some victims, that do not have ``open'' (publicly accessible) pictures but they have accessible liked pages. In this case \textit{S1} or \textit{S2} or \textit{S3} are strongly suggested.
%



\subsection{Experimental Results on Public Dataset}

\begin{figure}[h!]
	\subfigure[Average number of user IDs of friends found with respect to the number of HTTP requests, on Public Dataset. \label{fig:stratREQpub}]{\includegraphics[width=.47\textwidth]{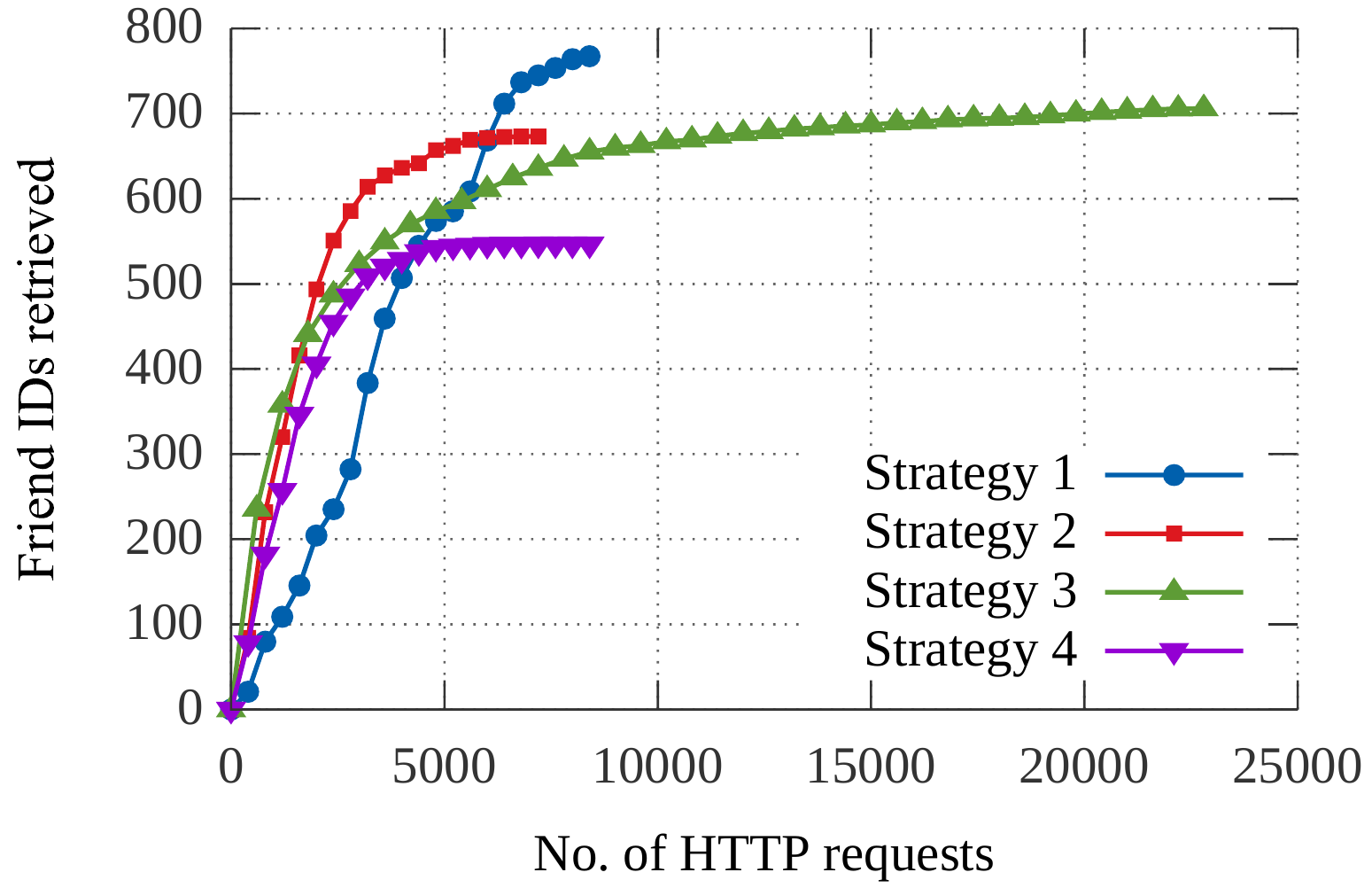}}
	\hfill
	\subfigure[Average number of user IDs of friends found against the time, on Public Dataset. \label{fig:stratTIMEpub}]{\includegraphics[width=.47\textwidth]{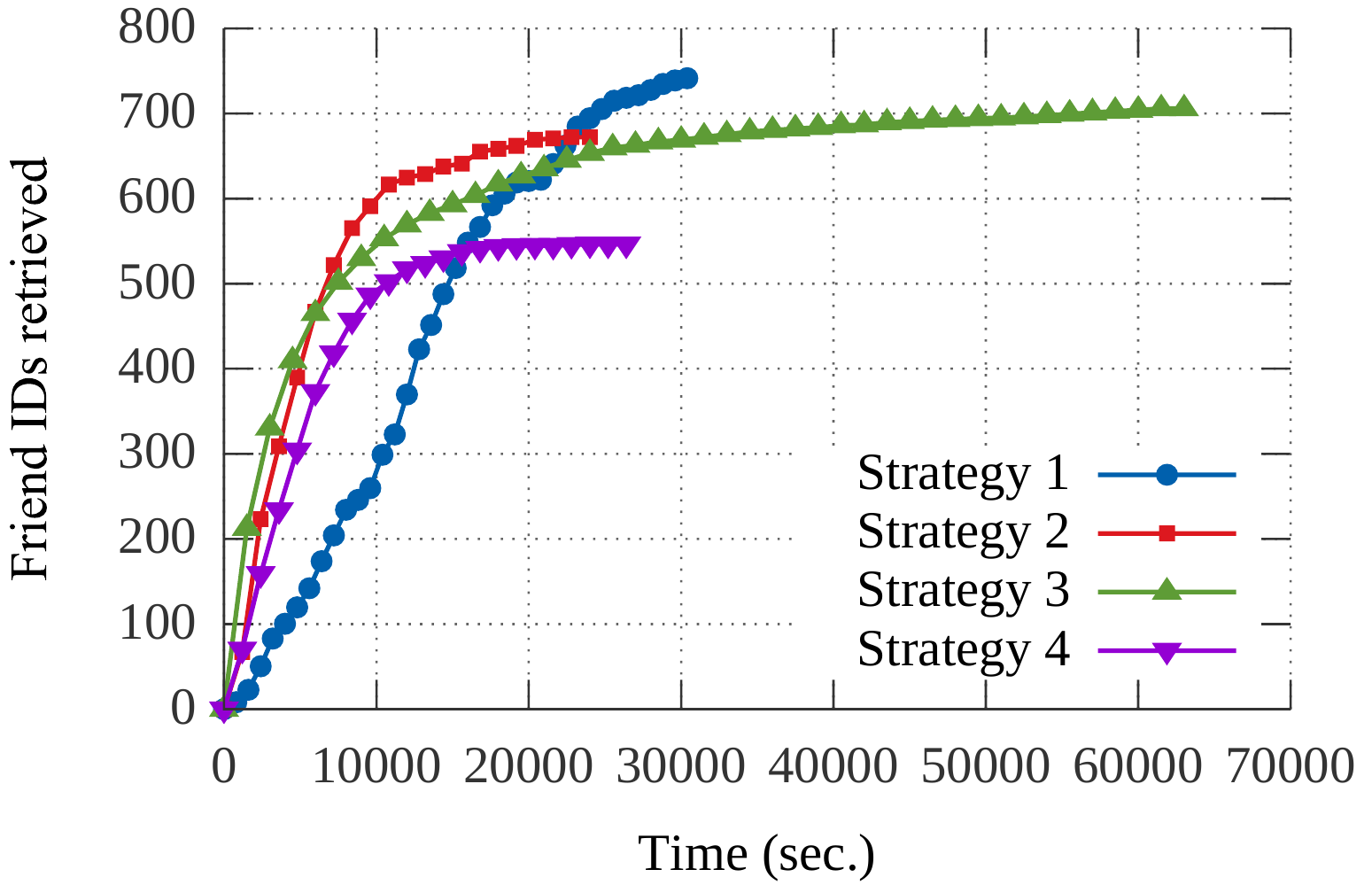}}
	\caption{Performance on the Public Dataset.}
	\label{fig:wrapper:public}
\end{figure}

The Public Dataset is composed by Facebook profiles that have a public profile.
The graphs in Fig.~\ref{fig:stratREQpub} and Fig.~\ref{fig:stratTIMEpub} present how our strategies become faster and more accurate on this dataset. As we expect, all the strategies work much better and find more user IDs of friends.
These results are perfectly in line with our assumption: people with a public profile have much more information available. Therefore, our strategies have more possibilities to find their friends.
It is possible to compare the curves reported in Fig.~\ref{fig:stratREQpub} and \ref{fig:stratTIMEpub} with those in Fig.~\ref{fig:stratREQ} and \ref{fig:stratTIME}. Experiment reported in Fig.~\ref{fig:stratREQpub} requires, on average, more ``HTTP requests'' than strategies applied to the dataset of Fig.~\ref{fig:stratREQ}. This is due to the fact that more information is available from the victim profile. This means that strategies applied to Public Dataset need more requests to retrieve more information than strategies on Mixed Dataset.
Also, the values of ``Friends IDs retrieved'' are higher, on average, in Fig.~\ref{fig:stratREQpub}. It means that the strategies applied to the Public Dataset (Fig.~\ref{fig:stratREQpub}) found more user IDs of friends compared to the same strategies on the Mixed Dataset (Fig.~\ref{fig:stratREQ}).\\
%
%
Generally speaking, the main difference between Mixed and Public Dataset is in terms of number of user IDs of friends found, which is much higher in the latter case. Although it seems meaningless to look for friends in the Public Dataset, we proposed these measures to prove the actual correctness of our approaches.

\subsection{Discussion}
%
We tested our approaches on a Mixed Dataset, composed by IDs of users that use privacy setting and IDs of users that have the Facebook profile publicly available.
On the Mixed Dataset composed from those user IDs that use privacy settings, the fastest strategy is \textit{S4}. It uses public pictures of a victim to retrieve IDs of friends. Strategies 1, 2 and 3 are slower than \textit{S4}, but when pictures are not publicly available, Strategies 1, 2 and 3 are the best alternative. 
In some cases we noticed that the strategies are able to retrieve up to 70\% of user IDs of friends of a victim.
On a Public Dataset all the strategies work better than on a Mixed Dataset: the number of IDs of friends found increases greatly. 
In this case, all four strategies retrieve a big amount of user IDs of friends. These results are in line with our thought.
%
Fig.~\ref{fig:stratREQ} and Fig.~\ref{fig:stratTIME}, from the Mixed Dataset, give us the main information. The biggest leakage of privacy is due to the public pictures. The percentage of user IDs of friends found from \textit{S4}, that exploits pictures, is 37.12\% with peaks of 70\%. Cover photos are always publicly available and it is not possible to make them private by Facebook privacy settings. This is the main reason that makes \textit{S4} to work better than the others and retrieves more user IDs of friends compared to \textit{S1} with 17.5\%, \textit{S2} with 17.4\% and \textit{S3} with 20.8\% of friends found.

On the other hand, we have to clarify the differences of percentages between \textit{S1}, 2 and~3. All these strategies work almost in the same way. The main motivation of these three different percentages is due to the FacePile plugin. For the same victim and the same like page, among the three strategies, the plugin shows a page with different set of candidate friends. This is motivated by the fact that our profiles are completely empty of information and do not have friends. Therefore, the Facebook engine, is not able to assign different ``distances'' among all the users, and thus it randomly shows candidate friends.

\section{Conclusion}
\label{conclusion}

The final aim of this work was to present a proof-of-concept approach that demonstrates a significant privacy issue on Facebook. Specifically, we exploited only tools publicly available in order to reveal information that the victim declared private.
The presented technique consists of four strategies, that rely on the above-mentioned tools, with the final aim of rebuilding the list of friends of a victim user (in the event that the victim declared such list private).
For our experiments we used two datasets of Facebook profiles, the Mixed Dataset and the Public Dataset. The first contains user IDs composed of profiles with different types of privacy settings and usually with information not publicly available; the latter contains user IDs profiles that are publicly available.
Since we used only public tools our evaluation shows how easy is to overcome the willingness of privacy of the users.
With our four strategies we can retrieve the friend list of every user with every type of privacy settings. Our percentages demonstrate a real lack of privacy in Facebook. We are now able to raise a real concern against Facebook. On the other hand, from our experiments, we hope to create awareness on Facebook users.

As a future work we want to analyze the possibility to rebuild not only a friend list, but the whole Facebook profile of a victim user.


\section*{Acknowledgments}

Mauro Conti is supported by a Marie Curie Fellowship funded by the European Commission under the agreement n. PCIG11-GA-2012-321980. This work has been partially supported by the TENACE PRIN Project 20103P34XC funded by the Italian MIUR, and by the Project ``Tackling Mobile Malware with Innovative Machine Learning Techniques'' funded by the University of Padua.

%

\sloppy
\bibliographystyle{splncs}

\bibliography{bibliografia.bib}

\end{document}